\documentclass{article}
\usepackage{spconf,amsmath,graphicx}
\usepackage{times}
\usepackage{epsfig}
\usepackage{graphicx}
\usepackage{amsmath}
\usepackage{amssymb}
\usepackage{array}
\usepackage{multirow}
\usepackage{booktabs}

\title{AnimePose: Multi-person 3D pose estimation and animation}
\name{Laxman Kumarapu$^{\dagger}$ \qquad  Prerana Mukherjee $^{\dagger}$  }
\address{ $^{\dagger}$ Dept. of Computer Science, Indian Institute of Information Technology, Sri City, India \\  $^{\dagger}$ \{laxman.k17, prerana.m\}@iiits.in  }
\begin{document}
%
\maketitle
\begin{abstract}
3D animation of humans in action is quite challenging as it involves
using a huge setup with several motion trackers all over the person’s
body to track the movements of every limb. This is time-
consuming and may cause the person discomfort in wearing
exoskeleton body suits with motion sensors. In this work, we present a trivial yet effective solution to generate 3D animation of multiple persons from a 2D video
using deep learning. Although significant improvement has been achieved recently in 3D human pose estimation, most of the prior works work well in case of single person pose estimation and multi-person pose estimation is still a challenging problem. In this work,
we firstly propose a supervised multi-person 3D pose estimation and animation framework namely \textit{AnimePose} for a given input RGB video sequence.
The pipeline of the proposed system consists of various modules: i) Person detection and segmentation, ii) Depth Map estimation, iii) Lifting 2D to 3D information for person localization iv) Person trajectory prediction and human pose tracking.
Our proposed system produces comparable results on previous state-of-the-art 3D multi-person pose estimation methods on publicly available datasets MuCo-3DHP and MuPoTS-3D datasets and it also outperforms previous state-of-the-art human pose tracking methods by a significant margin of 11.7\% performance gain on MOTA score on Posetrack 2018 dataset.

\end{abstract}


\section{Introduction}
\label{sec:intro}
The goal of 3D human pose estimation and animation is to localize semantic key points of a single or multiple human bodies in 3D space with texture and surface renditions. It is an essential technique for simulating the human behavior with the environment. 3D pose estimation and tracking is an essential step in action recognition tasks in video sequences. Recently,
many methods utilize deep convolutional neural networks (CNNs) and have achieved noticeable performance improvement on large-scale publicly
available 3D pose estimation datasets such as Human 3.6M dataset\cite{b35}.


The prior works in the context of human pose estimation can be categorized as i) 2D multi person pose estimation, ii) 3D single-person and iii) 3D multi-person pose estimation methods. There are two main
approaches in the 2D multi-person pose estimation. The first
one being top-down approaches\cite{b20}, which employ a human detector that estimates the bounding boxes of humans. Each detected human
area is cropped and fed into the pose estimation network
. The second one include the bottom-up approaches, which localize all human
body key points in an input image first, and then groups them into each person using some clustering techniques \cite{b21}.
Current 3D single person pose estimation methods can be categorized into
single- and two-stage approaches. The single-stage approach directly localizes the 3D body key points from the
input image \cite{b23}. The two-stage methods utilize the high accuracy of 2D human pose estimation. They initially localize
body key points in a 2D space and lift them to a 3D space. The second approach has achieved better results as the two-stage approach involves less complexity than the direct approach. Very few works address the problem of multi-person 3D pose estimation. This can be attributed to the scarcity of large datasets on multi-person 3D human pose estimation. However, there are significant contributions like . Moon \textit{et al.} \cite{b24} proposed a Camera Distance-aware top-down Approach for 3D Multi-person Pose Estimation from a Single RGB Image. Their proposed architectures consists of DetectNet, RootNet, and PoseNet. The DetectNet detects a human bounding box of each person in the input image. The RootNet
takes the cropped human image from the DetectNet and
localizes the root of the human R = (xR, yR, ZR), in
which xR and yR are pixel coordinates, and ZR is an absolute depth value. The same cropped human image is fed
to the PoseNet, which estimates the root-relative 3D pose.

In this work, we propose 3D multi-person pose estimation and animation framework namely \textit{AnimePose} for videos. First, we estimate 2D key points of a person in an input frame and also the depth map of that frame and then perform 3D pose estimation of a person by localizing the person in a 3D environment relative to the other persons with the help of the estimated depth map. Our approach not only works well for 3D multi-person pose estimation but it also outperforms previous human pose tracking methods as we have used human localization in 3D space using additional depth-map information and human trajectory estimation for tracking the humans in a video. 

\begin{figure*}
\centering
  \includegraphics[scale=0.2]{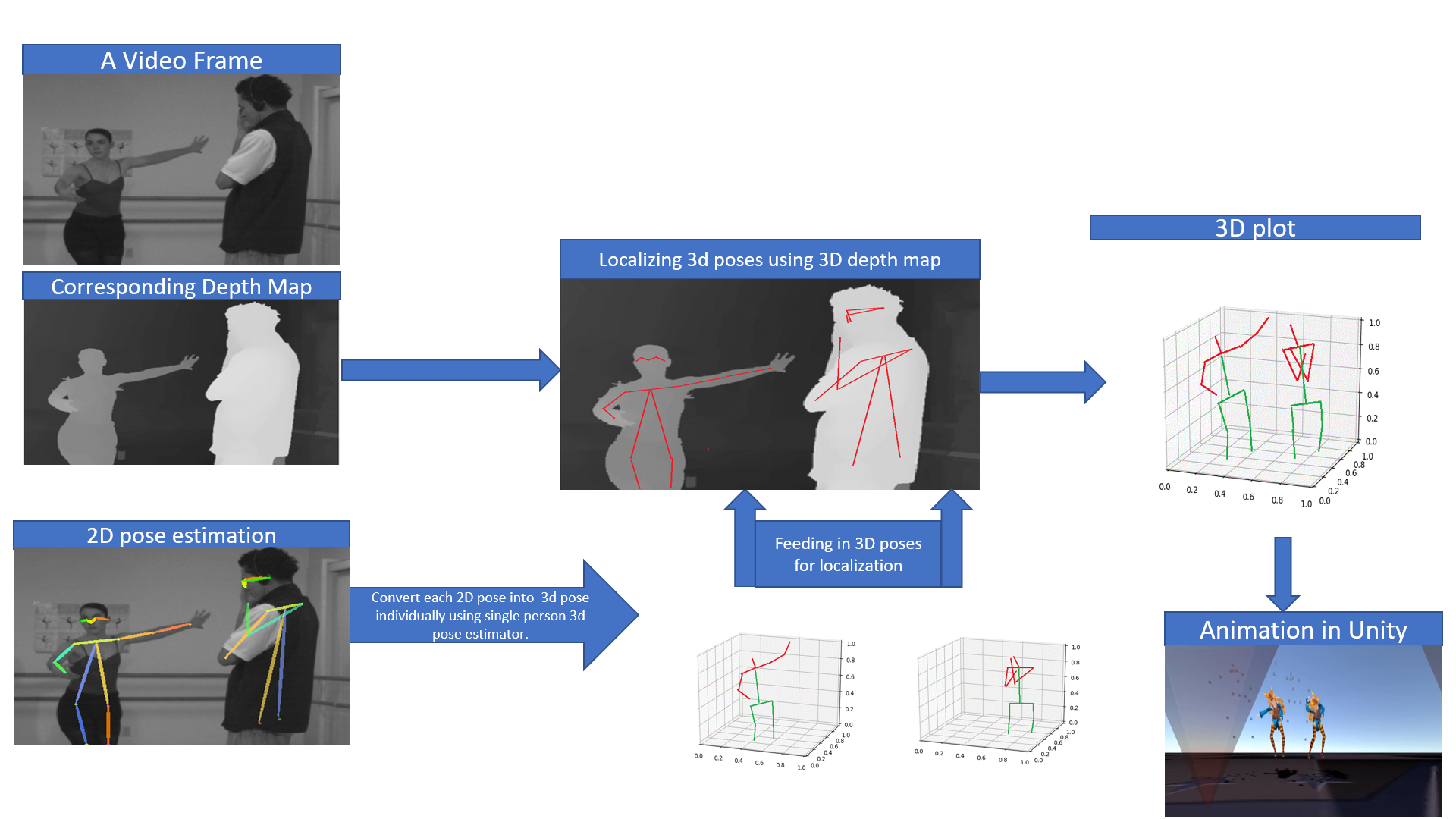}
  \caption{ Overall pipeline of the proposed framework for 3D multi-person pose estimation and animation in a frame-wise manner. The animation is generated in Unity3D environment.}
  \label{fig:fig1}
\end{figure*}

In view of above discussions the key contributions in our work can be summarized follows:
\begin{itemize}
\item To the best of author's knowledge, this is the first work to utilize depth map for localizing the 3D poses of multiple persons in 3D environment in order to ensure accurate relative 3D pose localization with respect to other objects in a 3D environment. 
\item We have also utilized person trajectory prediction mechanism and introduced a novel metric namely 3D IOU (generated with the help of depth maps) that helps us to overcome occlusion problem while predicting 3D poses in a video sequence and helps in better human tracking.
\item We have conducted exhaustive experiments on current 3D multi-person pose estimation methods and our proposed methodology  outperforms the current state of the art techniques by getting 82.1 3DPCKrel score on MuPoTS-3D dataset. Our pose tracking method using 3d human localization using depth map and person trajectory estimation outperforms the current state of the art by achieving total Multi-object tracking accuracy (MOTA) score of 60.1 \% on pose track 2018 dataset.
\end{itemize}


\section{Proposed Methodology}
\label{sec:methodology}

The goal of our proposed system is to predict 3D coordinates of key joints of multiple persons in a given video sequence. To solve this problem, we utilize a holistic top down approach that consists of i) Person Detection Network, Multi Person 2D pose estimation and
Depth Map estimation Network, ii) 2D to 3D pose uplifting Network and Relative localization of estimated 3D key joints with the help of Depth Map, and iii) Pose tracking module. Fig. \ref{fig:fig1} provides an overview of the proposed methodology.

\begin{figure}[htbp]
\centerline{\includegraphics[width=0.95\columnwidth]{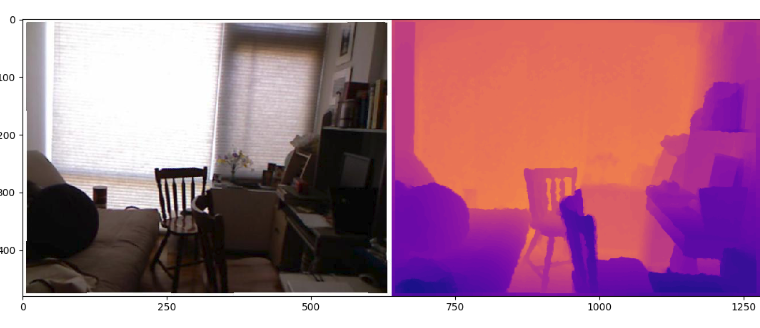}}
\caption{Input image(left) and the corresponding image depth map (right) .}
\label{fig:depth}
\end{figure}

\subsection{Multi-person 2D pose estimation and depth map generation}

We have utilized Hybrid task cascaded neural network with HR-Net(High Resolution Network) \cite{b25} as a backbone for person detection as it has provided State-of-the-art results over several object detection benchmarks. Next, we have utilized \cite{b22} Alpha pose detection network which is a bottom-up multi-person pose estimation approach. Recently, deep CNNs have shown great progress in solving depth estimation from a single Monocular image. Recent research such as Multi-Scale Local Planar Guidance \cite{b26} 
for Monocular Depth Estimation have shown promising results for tackling the depth estimation problem.
In the above-mentioned work they follow an encoder decoder scheme that reduces the feature map resolution to
H/8 then recovers back to the original resolution H. Here, Unlike the existing methods that
recovers back to the original resolution using simple nearest
neighbor up sampling layers and skip connections from encoding stages, the authors utilize novel local planar guidance (LPG)
layers which guide features to the full resolution with the
local planar assumption, and employ them together to get the final depth estimation. We have utilized this method for depth map generation of the input frame. Fig. \ref{fig:depth} shows the result obtained by the depth map generation \cite{b26}.

\subsection{2D to 3D multi person pose uplifting approach}
Significant progress has been achieved for 3D human
pose estimation from monocular images due to the availability of large-scale dataset \cite{b1} and the powerful deep learning frameworks.
Two-stage approaches first estimate 2D poses and then
lift 2D poses to 3D poses \cite{b1}. 
These approaches usually generalize better on images in
the wild, since the first stage can benefit from the state-of-the-art 2D pose estimators, which can be trained on images in the wild. The second stage usually regresses the
3D locations from the 2D predictions. For example, Martinez \textit{et al.} \cite{b11} proposed a simple fully connected residual
networks to directly regress 3D coordinates from 2D coordinates. But the problem with this two stage approach is that it works greatly for single person pose estimation as that network takes one pose at a time and convert it into 3D poses. In the process of doing this, we lose the relative position of one person with respect to others and thus it fails in multi person 3D pose estimation. In order to cater solution to this, we preserve the relative position by proposing the localization in 3D environment utilizing depth information. First, we take the pose and its corresponding 2D bounding box co-ordinates then we uplift the 2D poses to 3D poses individually. Using the depth map, bounding box co-ordinates and segmentation mask we localize it in a 3D environment as we can infer 3D bounding box of a person. Due to this combination of depth map and bounding boxes we are able to preserve the relative position between the poses. We then project it into the 3D environment using appropriate scaling and here the relative positions are preserved.

In order to get 3D bounding box, we take 2D bounding box and a corresponding 2d segmentation mask of the person. We then look for the points
$\mathrm{Z}_\mathrm{min}$ that is closer to the camera  and $\mathrm{Z}_\mathrm{max}$ that is far away from camera within the segmentation mask regions of the person and if 2D bounding box co-ordinates are \{($\mathrm{X}_\mathrm{min}$ ,$\mathrm{Y}_\mathrm{min}$) ,($\mathrm{X}_\mathrm{max}$ ,$\mathrm{Y}_\mathrm{max}$)\} then the corresponding 3D coordinates of the bounding box are estimated as,
\{($\mathrm{X}_\mathrm{min}$ ,$\mathrm{Y}_\mathrm{min}$,$\mathrm{Z}_\mathrm{min}$),
($\mathrm{X}_\mathrm{max}$ ,$\mathrm{Y}_\mathrm{max}$,$\mathrm{Z}_\mathrm{min}$),
($\mathrm{X}_\mathrm{min}$ ,$\mathrm{Y}_\mathrm{min}$,$\mathrm{Z}_\mathrm{max}$),
($\mathrm{X}_\mathrm{max}$ ,$\mathrm{Y}_\mathrm{max}$,$\mathrm{Z}_\mathrm{max}$)\}
respectively. Fig. \ref{fig:3dboxes} shows the 3D localization for multiple persons in a given frame.

\subsection{Multi person pose tracking}
In order to extend the approach to the entire video sequence, we need to keep track of individual poses. Recent work like
efficient multi-person 2D Pose Tracking with
Recurrent Spatio-Temporal Affinity Fields \cite{b32} has achieved promising results on multi person pose tracking.
But these methods are not robust to occlusion and overlapping conditions. To circumvent these problems, along with the procedure mentioned in \cite{b32}, we introduce a novel solution that utilizes depth map information and temporal  person trajectory estimation.
\begin{figure}[htbp]
\centerline{\includegraphics[width=0.95\columnwidth]{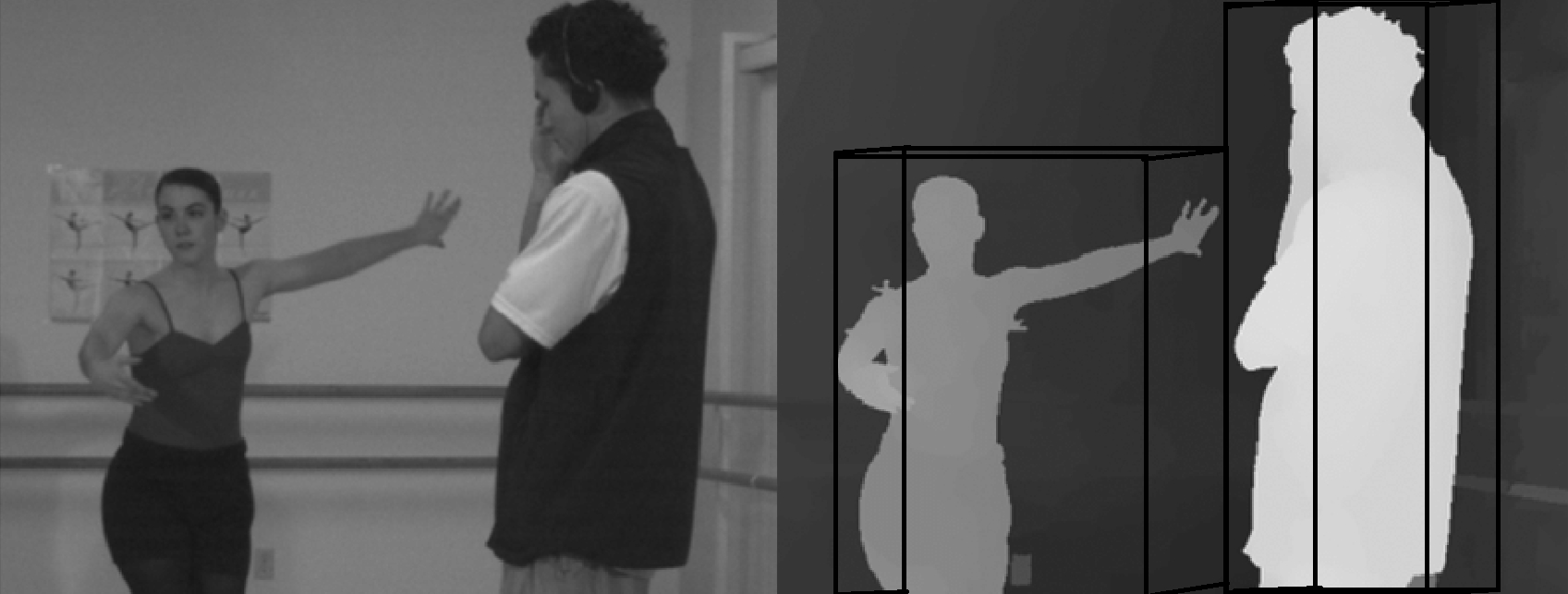}}
\caption{Input frame (left) and corresponding 3D localization of the persons done with the help of estimated depth map (right).}
\label{fig:3dboxes}
\end{figure}

\begin{table*}[t]
\centering
\caption{Sequence-wise 3DPCKrel \cite{b28} comparison with state-of-the-art methods on the MuPoTS-3D dataset}
 \setlength{\tabcolsep}{2pt}
\begin{tabular}{||c c c c c c c c c c c c c c c c c c c c c c ||}
 
  \hline\hline
 
 Methods & S1 & S2 & S3 & S4 & S5 & S6 & S7 & S8 & S9 & S10 & S11 & S12 & S13 & S14 & S15 & S16 & S17 & S18 & S19 & S20 & Avg \\ [0.5ex] 
 \hline\hline
 Rogez \cite{b27} & 67.7& 49.8 & 53.4 & 59.1 & 67.5 & 22.8 & 43.7 & 49.9 & 31.1 & 78.1 & 50.2 & 51.0 & 51.6 & 49.3 & 56.2 & 66.5 & 65.2 & 62.9 & 66.1 & 59.1 & 53.8\\ 
 \hline
 Mehta \cite{b28} & 81.0& 60.9 & 64.9 & 63.0 & 69.1 & 30.3 & 65.0 & 59.6 & 64.1 & 83.9 & 68.0 & 68.6 & 62.3 & 59.2 & 70.1 & 80.0 & 79.6 & 67.3 & 66.6 & 67.2 & 66.0\\ 
\hline
 3D MPPE \cite{b24} & \textbf{94.4}& 77.5 & \textbf{79.0} & \textbf{81.9} & 85.3 & 72.8 & 81.9 & \textbf{75.7} & \textbf{90.2} & 90.4 & 79.2 & \textbf{79.9} & 75.1 & \textbf{72.7} & 81.1 & 89.9 & \textbf{89.6} & 81.8 & 81.7 & \textbf{76.2} & 81.8\\ 

 \hline
 \textbf{Ours} & 94.2& \textbf{77.8} & 78.8 & 81.7 & \textbf{85.5} & \textbf{73.1} & \textbf{82.0} & 75.5 & 90.0 & \textbf{90.8} & \textbf{79.7} & 79.7 & \textbf{75.4} & 72.5 & \textbf{81.3} & \textbf{90.1} & 89.4 & \textbf{82.1} & \textbf{81.9} & 76.0 & \textbf{82.1}\\ 
 \hline
 
\end{tabular}
\par
\bigskip
\label{tab:seqmupots}
\end{table*}

To track a pose, on the top of 2D pose prediction we propose 3D IOU (Intersection Over Union) metric instead of traditional 2D IOU which can be inferred with the help of bounding box co-ordinates and depth map as shown in Fig. \ref{fig:3dboxes}. In order to overcome occlusion problem, if the entire pose of the person gets missing whereas it is present in previous frames, we perform temporal person trajectory estimation using  Scene-LSTM: A Model for Human Trajectory Prediction \cite{b36} to fill the missing tracks in the occluded frames in a video. The way we predict the missing poses is that we take a set of previous frames and plot the trajectory of each key points and estimate the position where the point will be in the future frames and we utilize depth in addition to it so that we may overcome the overlapping issue of 2 poses in a 2D frame. We have trained our Scene-LSTM on 50\% of PoseTrack dataset. Given a 3D bounding boxes $B_1$ in previous frame and its tracked 3D bounding box $B_2$ in next frame, the formulation of 3D IOU ($IOU_{3D}$) is as follows, 
\begin{equation}
    IOU_{3D}=\frac{V_{B_1} \cap V_{B_2}}{V_{B_1} \cup V_{B_2}}
\end{equation}
where $V_{B_1}$ and $V_{B_2}$ denote the volumes of the two 3D bounding boxes in two consecutive frames.

\section{Experiments and Results}\label{sec:results}

\subsection{Dataset and Evaluation Metric}

\textbf{ MuCo-3DHP and MuPoTS-3D datasets:} These are
the 3D multi-person pose estimation datasets proposed by
Mehta \textit{et al.} \cite{b28}. The training set MuCo-3DHP is generated by compositing the existing MPI-INF-3DHP 3D
single-person pose estimation dataset \cite{b34}. The test set
MuPoTS-3D dataset was captured at outdoors and it includes 20 real-world scenes with groundtruth 3D poses for
up to three subjects. The groundtruth is obtained with a
multi-view marker-less motion capture system. For evaluation, a 3D percentage of correct keypoints (3DPCKrel) and
area under 3DPCK curve from various thresholds (AUCrel)
is used after root alignment with groundtruth.

\begin{table}
\caption{Joint-wise 3DPCKrel \cite{b28} comparison with state of-the-art methods on the MuPoTS-3D dataset.}
\setlength{\tabcolsep}{1pt}
 \begin{tabular}{||c c c c c c c c c c||} 
 
 \hline\hline
 Methods & Hd. & Nck. & Sho. & Elb.& Wri.& Hip. &Kn. & Ank. & Avg \\ [0.5ex] 
 \hline\hline
 Rogez \cite{b27} & 49.4 & 67.4 & 57.1 & 51.4 & 41.3 & 84.6 &56.3 & 36.3 & 53.8 \\ 
 \hline
 Mehta \cite{b28} & 62.1 & 81.2 & 77.9 &57.7 & 47.2 & 97.3 & 66.3 & 47.6 & 66.0 \\
 \hline
 3D MPPE \cite{b24} & 79.1 & 92.6 & \textbf{85.1} & \textbf{79.4} & \textbf{67.0} & 96.6 & 85.7 & 73.1 & 81.8\\
 \hline
 \textbf{Ours} & \textbf{79.3} & \textbf{92.8} & 84.9 & 79.3 & 66.8 & \textbf{97.4}& \textbf{86.1} & \textbf{73.3} & \textbf{82.1} \\ 
 \hline
\end{tabular}
\par
\bigskip
\label{tab:tabMupots}
\end{table}

\textbf{ Pose Track 2018 dataset:}
PoseTrack is a large-scale benchmark for human pose estimation and articulated tracking in video. They provide a publicly available training and validation set as well as an evaluation server for benchmarking on a held-out test set. For evaluation, they use MOTA (Multi object tracking Accuracy) on the poses in a video. 



\begin{table}
\caption{Pose Track 2018 validation results}
\setlength{\tabcolsep}{1pt}
 \begin{tabular}{||c c c c c ||} 
 
  \hline\hline
 
 Method & Wrist-AP & Ankles-AP & mAP & MOTA \\ [0.5ex] 
 \hline\hline
 PoseTrack \cite{b29} & 54.3 & 49.2 & 59.4 & 48.4 \\ 
 \hline
 BUTD \cite{b30} & 52.9 & 42.6 & 59.1 & 50.6 \\
 \hline
 PoseFlow \cite{b31} & 59.0 & 57.9 & 63.0 & 51.0 \\
 \hline
 JointFlow \cite{b32} & 53.1 & 50.4 & 63.3 & 53.1 \\
 \hline
 Temporal affinity fields\cite{b33} & 65.0 & \textbf{60.7} & \textbf{70.3} & 53.8 \\  
 \hline
 \textbf{Ours} & \textbf{65.2} & 60.2 & 70.2 & \textbf{60.1} \\ 
 \hline
\end{tabular}
\par
\bigskip
\label{tab:posetrack}
\end{table}

\subsection{Quantitative and Qualitative Results}
Table \ref{tab:seqmupots} and \ref{tab:tabMupots} gives the sequence-wise and joint-wise 3DPCKrel score comparative analysis on MuPoTS-3D dataset respectively. Our proposed approach has produced comparable results with State-of-the-Art techniques on MUPoTS-3D dataset by achieving an average of 82.1 on 3DPCKrel evaluation criteria. The results on the Pose Track dataset is provided in Table \ref{tab:posetrack}. Our approach performs better than current  State-of-the-Art pose tracking techniques and achieves a MOTA of 60.1\% on Pose Track 2018 Validation dataset.

\begin{figure}[htbp]
\centerline{\includegraphics[width=0.95\columnwidth]{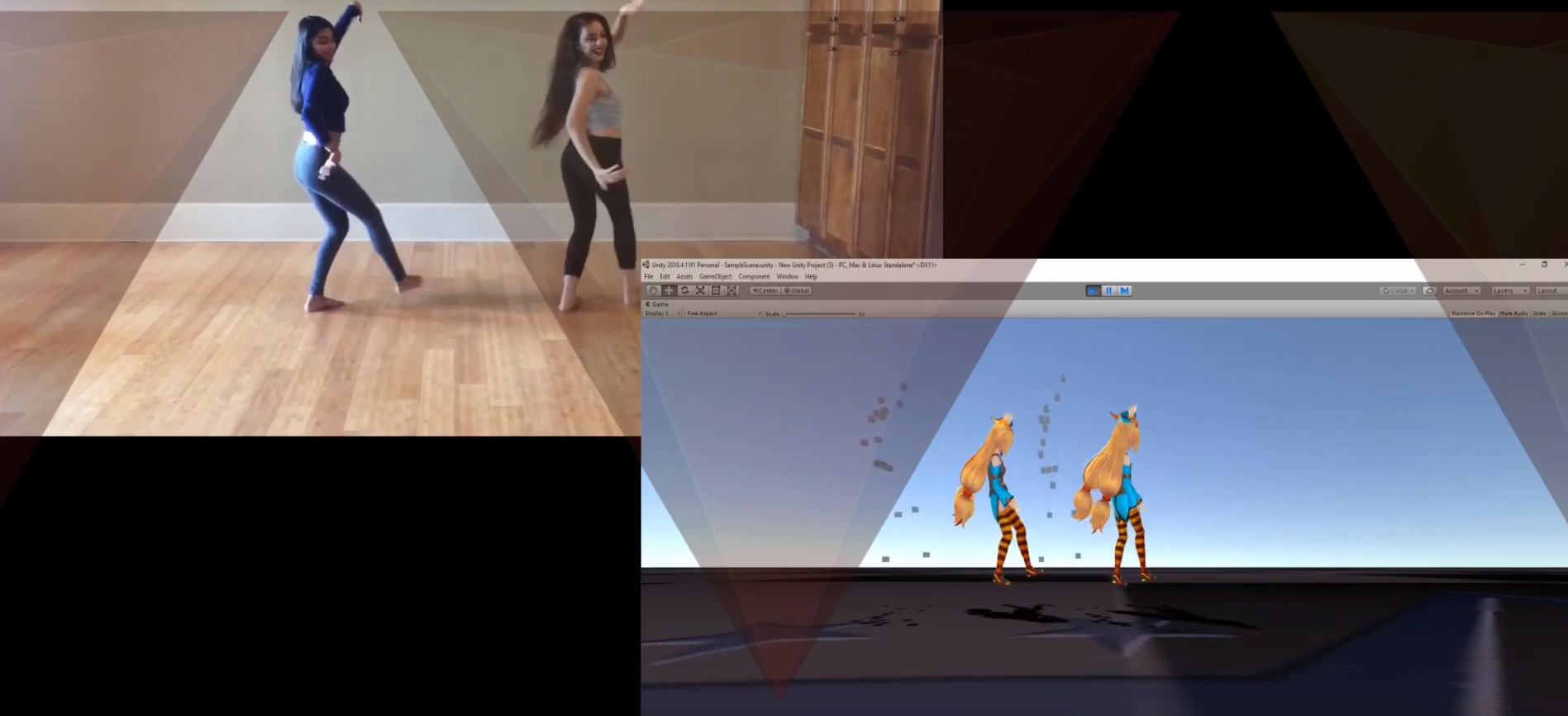}}
\caption{Multi-person person 2D RGB frame and its corresponding 3D animation.}
\label{fig:animation}
\end{figure}

We have provided results of our 3D multi person pose estimation and multi person animation on an input frame  using Unity3D in Fig. \ref{fig:animation}. We are able to effectively localize the persons in 3D and render animation.

\section{Conclusion}
\label{sec:conclusion}
We proposed a novel and holistic framework for 3D multi-person pose estimation  and animation in a frame-wise manner in videos.
We are effectively able to localize the 3D relative poses in a 3D environment due to the use of depth information. Further, we proposed a novel framework for pose tracking in which along with recurrent spatio-temporal affinity fields we used depth information to calculate a novel metric 3D IOU instead of traditional 2D IOU and utilized multi person trajectory estimation to continue tracking a person in occluded frames. The proposed method produced comparable results to previous 3D multi-person pose
estimation methods without any ground truth information while other methods
utilize it during inference.


\end{document}